\documentclass[twocolumn,aps,prl,superscriptaddress,showpacs]{revtex4}
\usepackage{amsmath,bm}%
\usepackage{graphicx}%

\begin{document}

\draft
\title{ Total Reaction Cross Section in an Isospin-Dependent Quantum Molecular Dynamics (IDQMD) Model }
\thanks{ Supported by the Major State Basic Research Development Program of China under Grant No TG 2000077404. }

\author{ Y. B. Wei}
\affiliation{Shanghai Institute of Nuclear Research, Chinese Academy of Scienecs, P.O. Box 800-204,
Shanghai 201800, China}
\author{X. Z. Cai}
\affiliation{Shanghai Institute of Nuclear Research, Chinese Academy of Scienecs, P.O. Box 800-204,
Shanghai 201800, China}   
\author{ W. Q. Shen}
  \affiliation{Shanghai Institute of Nuclear Research, Chinese Academy of Scienecs, P.O. Box 800-204,
Shanghai 201800, China}   
\affiliation{   CCAST (World Laboratory), PO Box 8730, Beijing 100080, China}
  \affiliation{  Department of Physics, Ningbo University, Ningbo 315211, China}
\author{Y. G. Ma } 
  \affiliation{Shanghai Institute of Nuclear Research, Chinese Academy of Scienecs, P.O. Box 800-204,
Shanghai 201800, China}   
   \affiliation{   CCAST (World Laboratory), PO Box 8730, Beijing 100080, China}
\author{H. Y. Zhang}
\author{C. Zhong}
\author{W. Guo }
\author{J. G. Chen}
\author{G. L. Ma}
\author{K. Wang} 
 \affiliation{Shanghai Institute of Nuclear Research, Chinese Academy of Scienecs, P.O. Box 800-204,
Shanghai 201800, China}   
 \date{\today}

\begin{abstract}

The isospin-dependent quantum molecular dynamics (IDQMD) model is used to study the total reaction cross 
section $\sigma_R$. 
The energy-dependent Pauli volumes of neutrons and protons have been discussed and introduced into the IDQMD 
calculation to replace the widely used energy-independent Pauli volumes. The modified IDQMD calculation can 
reproduce the experimental $\sigma_R$ well for both stable and exotic nuclei induced reactions. 
Comparisons of the calculated $\sigma_R$ 
induced by $^{11}Li$ with different initial density distributions have been performed. It is shown that 
the calculation by using 
the experimentally deduced density distribution with a long tail can fit the 
experimental excitation function better than 
that by using the Skyrme-Hartree-Fock calculated density without long tails. 
It is also found that $\sigma_R$ at high energy is 
sensitive to the long tail of density distribution. 

\end{abstract}


\maketitle

The total reaction cross section $\sigma_R$ has been extensively studied theoretically and 
experimentally.[1-10] There are two kinds of theoretical models to calculate $\sigma_R$. The first is the 
low energy theory based on the interaction potential. Such models are not successful for the 
reaction beyond 10-15 MeV/nucleon above the Coulomb barrier. The second is the high energy 
microscopic Glauber theory based on the individual nucleon-nucleon collisions in the overlap 
volume of the projectile and target. However, the roles of mean field and medium effect are 
difficult to be discussed in the Glauber-type model. In recent years, Ma {\it et\ al}. developed a new 
method to study $\sigma_R$ with helps of transport theory and Glauber model.[5] Originally, the BUU 
model [11] was taken as a tool to investigate $\sigma_R$.[5] Later, the quantum molecular dynamics 
(QMD) model was applied to study $\sigma_R$ [13] in the same spirit as Ref. [5]. The reaction dynamics 
in transport theory at intermediate energy is mainly governed by the mean field, two-body 
collisions, and Pauli blocking. To investigate the isospin effects, the above three dynamical 
ingredients should include properly isospin degrees of freedom to obtain an isospin-dependent 
quantum molecular dynamics (IDQMD). It is also important that the samples of neutrons and 
protons in the phase space should be treated separately in the initialization of projectile and 
target nuclei. In this Letter, the IDQMD model is introduced to calculate $\sigma_R$. This model 
incorporates the isospin-dependence of mean-field, nucleon-nucleon cross section, and Pauli 
blocking. It has been widely used to study the multi-fragmentation and the collective flow. In the 
IDQMD model, neutrons and protons are distinguishable in the initialization.

The density distributions for the initial projectile and target nuclei are determined from the 
Skyrme-Hartree-Fock (SHF) calculation [17] with parameter set SKM. The stability of the 
propagation of the initialized nuclei has been checked in details and can last at least 200fm/c 
according to the evolutions of the average binding energies and the root mean square radii of the 
initialized nuclei. 

In the IDQMD model, once the distance between the two nucleons is less than rnn, the 
collision may occur, where rnn is defined as
\begin{equation}
r_{nn} = \sqrt{\sigma_{nn}(\sqrt{s})/\pi} \nonumber
\end{equation}
with $\sigma_{nn}(\sqrt{s})$ representing the nucleon-nucleon reaction cross section and  $\sqrt{s}$
 representing the 
nucleon energy in the center of mass system. Whenever a collision occurs, the final momenta of 
the scattering nucleons can be easily obtained as a consequence of momentum conservation and 
the coordinates are updated in terms of newtonian equation (classical trajectory). The 
six-dimensional phase space radius of one nucleon is the product of its $\Delta$P and $\Delta$R. Then we 
calculate and check the phase spaces around the final states of the scattering nucleons. Thus it is 
easy to determine the probabilities ($P_1$ and $P_2$) for each of the two scattering nucleons that their 
final phase spaces are already occupied by other nucleons by comparing with the defined Pauli volue,
i.e. 4$(\Delta P \Delta R)^3/3$, in six-dimensional phase space, where $\Delta R$ is the minimum
radius which is allowed to be occupied by itself in corrdinate space and the $\Delta P$ is the 
same but in momentum space. The collision is then blocked with a 
probability
\begin{equation}
P_{block} = 1-[1-min(P_1,1)][1-min(P_2,1)]. \nonumber
\end{equation}
Correspondingly, the collision is allowed with probability $(1-P_{block})$. Whenever a collision is 
blocked, the momenta of the scattering nucleons will be replaced by the values they are prior to 
scattering. More details could be found in Refs. [12,15].

The following formula of in-medium reaction cross section is used in the present IDQMD 
calculation, 
 \begin{eqnarray}
 \sigma_{nn} = (13.73-15.04\beta ^{-1} + 8.76 \beta ^{-2} + 68.67 \beta ^{-4}) \nonumber \\
  \cdot \frac{1.0 + 7.772 E_{lab}^{0.006} \rho ^{1.48}} {1.0 + 18.0 \rho ^{1.46}} \nonumber\\
\sigma_{np} = (-70.67 -18.18 \beta ^{-1} + 25.26 \beta ^{-2} + 113.85 \beta)\nonumber\\
\cdot \frac{1.0 + 20.88 E_{lab}^{0.04} \rho ^{2.02}} {1.0 + 35.86 \rho ^{1.90}}\nonumber\\
\beta = \sqrt{1-\frac{1.0}{\gamma ^2}}, \gamma = \frac{E_{lab}}{931.5} + 1
\end{eqnarray}
This formula includes both effects of incident energy Elab and nucleon matter density $\rho$. It shows 
that the medium effect is important at intermediate energies and becomes smaller at higher 
energies but does not vanish.[18, 19].

The cross section $\sigma_R$ can be written as [5] 
\begin{equation} 
 \sigma_R = 2\pi \int {b[1-T(b)]db} = 2\pi \int {b[1-exp(-N)] db}
\end{equation}
where the transport function T(b) can be obtained from the average n-n collision number N as a 
function of the impact parameter b. More details could be found in Ref. [5]. 

In the previous IDQMD model, the volume occupied by nucleon in the projectile and target 
(Pauli volume) was a constant ($\hbar^3/2$). This constant could be deduced from the lowest limit of 
the uncertain relationship between the momentum and the coordinate.[14] Recently it was found 
that the Pauli volume is sensitive to the incident energy and should not be a constant 
qualitatively.[20] Calculation of the average collision number $N$ of the $^{12}C$ +$^{12}C$ system at high 
incident energy shows that the average collision number still has an uptrend after the evolution 
time of 200fm/c. However, other studies indicated that the average collision number has been 
saturated after 50fm/c at high incident energy.[13] This indicates that the invariable Pauli 
volume used in the previous IDQMD model may be not suitable and the Pauli volume may be 
energy-dependent. From formula (2) we know that $\sigma_R$ can be decided by the average collision 
number $N$. It is also clear that N is sensitive to the probability of the Pauli blocking in the 
collisions. Thus, the energy dependence of the Pauli volume becomes very important in the 
calculation of $\sigma_R$. Up to now, there has not yet been parameterized formula to describe the 
energy-dependent Pauli volume. 

We calculate the cross section $\sigma_R$ of the $^{12}C$ + $^{12}C$ reaction system in an IDQMD framework 
in a wide energy range as shown in Fig. 1(a). Here soft EOS and in-medium $\sigma_{nn}$ are used. The 
initialized density distribution of $^{12}C$ comes from the SHF calculation. The solid line with open 
circles shows the calculated results with the constant Pauli volume. It is obvious that they can 
not reproduce the experimental data. With adjusted Pauli volume values in the IDQMD 
calculation we fit the experimental data at different energies [see the dashed line with open 
squares in Fig.1(a)]. From these fits, the Pauli volume values were obtained in a wide energy 
range as shown in Fig. 1(b). Obviously, the Pauli radius R increases generally at intermediate 
energy and becomes saturated above 400 MeV/u. By using this energy-dependent Pauli volume, 
it is found that the average collision number N will be a constant after 50fm/c at high energy and 
after 100fm/c at intermediate energy, which is consistent with the other studies.[13] 

\begin{figure}
\includegraphics[scale=0.35]{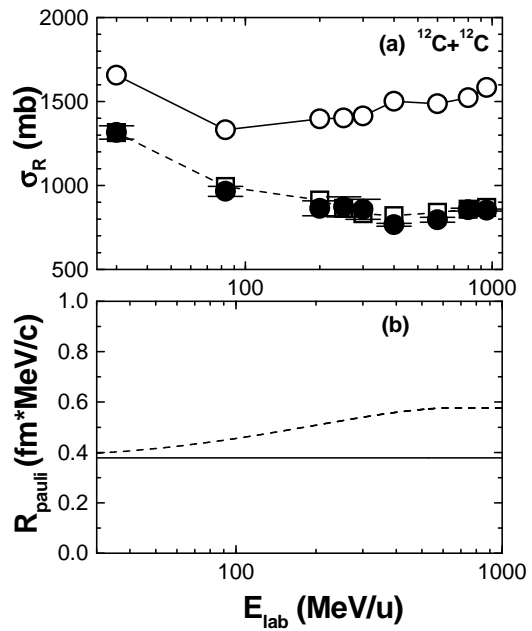}
\caption{\footnotesize (a) Total reaction cross section $\sigma_R$
 of $^{12}C$ on a $^{12}C$ target. The solid line with 
open circles shows the calculated 
results with the constant Pauli volume values, the dashed line with open squares represents 
the calculated results with the 
changeable Pauli volume values, the closed circles are experimental data from Ref. [4]. (b) 
Solid line is the constant Pauli radius 
R0 taken in the previous IDQMD model and the dashed line shows the energy-dependent Pauli
 radius R in this work.}
\label{fig1}
\end{figure}

In order to test the modified IDQMD model, we use it to study $\sigma_R$ for other reaction systems. 
In Fig.2, the solid line shows the experimental matter density distribution with a long tail and the 
dashed line shows the one calculated with the SHF density distribution. The experimental 
density distribution of $^{11}Li$ is introduced into the IDQMD model to replace the SHF one. Figure 
3(a) shows the calculated results of $\sigma_R$for the $^{11}Li$ + $^{12}C$ reaction system with different density 
distributions as shown in Fig.2. Clearly, the calculation with the experimental density 
distribution of $^{11}Li$ can fit the experimental $\sigma_R$ better. For the reaction induced by halo 
nucleus$^{11}Li$, the results with the SHF density distribution are lower than the experimental ones at 
high incident energy for about a few percent since the SHF calculation of $^{11}Li$ does not give the 
long tail as experimental one, which is expected to play an important role in the calculation of $\sigma_R$. 
This indicates that the density distribution calculated with the SHF density distribution is not 
appropriate to those halo nuclei [21] in the IDQMD calculation. It is obvious that the calculated 
$\sigma_R$ at several hundreds of MeV/u is more sensitive to the long-tail density distribution. Since the 
central densities are also different between them, more conclusions about the density effect on 
$\sigma_R$ can be obtained by further studies. The research along this line is in progress.

 \begin{figure}
\includegraphics[scale=0.35]{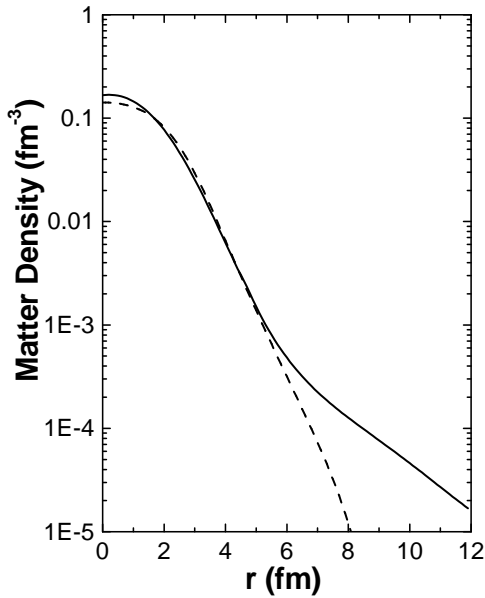}
\caption{\footnotesize Experimental density distribution 
(solid) of $^{11}Li$ ($\rho_1$) [21] and the calculated one (dashed) with the 
Skyrme-Hartree-Fock density distribution $\rho_0$.}
\label{fig2}
\end{figure}

 \begin{figure}
\includegraphics[scale=0.35]{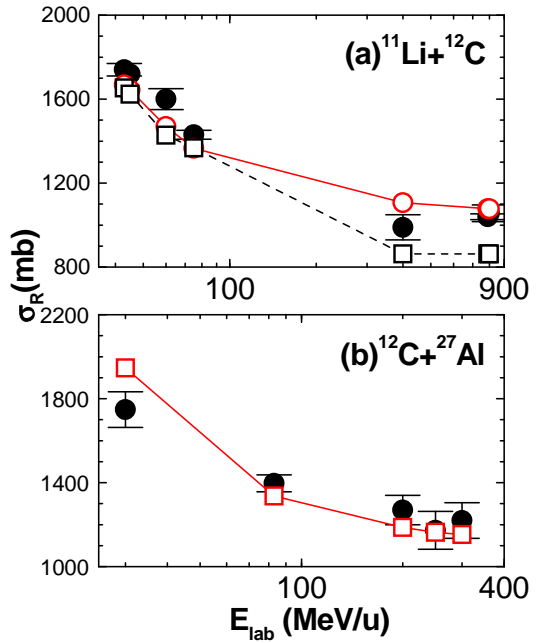}
\caption{\footnotesize (a) Calculation of $\sigma_R$ 
for $^{11}Li$ on $^{12}C$ target. The dashed line with 
open squares shows the calculated results with $\rho_0$ and 
the solid line with open circles represents the calculated results with $\rho_1$. 
(b) The solid line with open square shows the energy 
dependence of $\sigma_R$for the $^{12}C$ + $^{27}$Al. 
The closed circles are the experimental data from Refs. [4,22].}
\label{fig3}
\end{figure}

Figure 3(b) shows the calculated results of excitation function of $\sigma_R$for the $^{12}C$ + $^{27}Al$ 
reaction system. In Fig. 3, all the filled circles represent the experimental data.[4,22] The solid 
line with open squares shows the calculated results with the IDQMD, where the initialized 
density distribution of $^{27}Al$ and $^{12}C$ are calculated by using the SHF model. Figure 3 shows that 
the IDQMD calculation with the energy-dependent Pauli volume and the density distribution of 
the SHF model gives a good trend of excitation function of $\sigma_R$. For reaction induced by stable 
nuclei, the calculated results can fit the experimental values well in a wide energy range. With 
the experimental density distribution the IDQMD model can also give the fine results of $\sigma_R$
induced by halo nuclei. 

Figure 4 gives the calculation of $\sigma_R$of Li isotopes on the $^{12}C$ target at 790MeV/u. The 
density distribution of $^{11}Li$ is from the experiment and the others are calculated with the SHF 
model. It shows that $\sigma_R$ varies smoothly with mass number up to 9Li. There is a sudden increase 
of $\sigma_R$ between 9Li and $^{11}Li$ which is corresponding to the halo structure in $^{11}Li$. Our calculation 
results reproduce both the experimental values and the sudden change between 9Li and $^{11}Li$ 
fairly well. 
                 
 \begin{figure}
\includegraphics[scale=0.35]{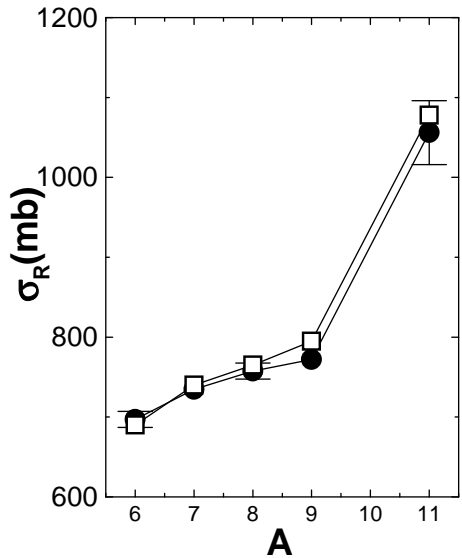}
\caption{\footnotesize Total reaction cross section $\sigma_R$ of 
Li isotopes on a $^{12}C$ target at 790 MeV/u. 
The solid line with open squares 
shows the calculated results and the solid line with filled circles
 represents the experimental data taken from Refs. [10,23].}
\label{fig4}
\end{figure}

In conclusion, the IDQMD model has been introduced to study the total reaction cross section $\sigma_R$ by 
using the energye-dependent Pauli volume which is deduced from the excitation function of $\sigma_R$
of $^{12}C$+$^{12}C$. The calculated results can reproduce the experimental $\sigma_R$ quite well. It is interesting 
to investigate further the energy and isospin effects of Pauli volume in collision. For halo nuclei, 
the calculated results by using experimental density distribution are better than that by using the 
SHF density distribution. It is suggested that the long tail of the density distribution plays an 
important role for the halo nuclei and $\sigma_R$ at high energy is sensitive to the long tail distribution of 
halo nuclei. Since the IDQMD model incorporates the isospin-dependences of mean-field, 
nucleon-nucleon cross section, and Pauli blocking, it is also interesting to study each 
isospin-effect on $\sigma_R$ and other physical quantities. In the IDQMD calculation, the clusters can be 
judged by the relative momenta and coordinates with an isospin-dependent modified 
coalescence model and the momentum distribution of the projectile fragments can be obtained 
easily. Thus, the IDQMD model can be used to study the total reaction cross section, 
fragmentation cross sections and momentum distribution of fragment of halo nuclei 
simultaneously. It can be used to give a more comprehensive and reliable criterion of halo 
structure, which is of great significance. These works are in progress.

One of the authors (Y. B. Wei) gratefully acknowledges Dr. F. S. Zhang for his 
enthusiastic help in personal discussion.

{}

\end{document}